\documentclass[aps,prc,groupedaddress,twocolumn,showpacs]{revtex4-1}
\usepackage{graphicx}
\usepackage{dcolumn}

\begin{document}

\title{Investigation of $\alpha$ induced reactions on $^{130}$Ba and $^{132}$Ba 
and their importance for the synthesis of heavy $p$ nuclei}

\author{Z. Hal\'asz}
\author{Gy. Gy\"urky}\email[corresponding author, ]{gyurky@atomki.hu}
\author{J. Farkas}
\author{Zs. F\"ul\"op}
\author{T. Sz\"ucs}
\author{E. Somorjai}
\affiliation{Institute of Nuclear Research (ATOMKI), H-4001 Debrecen, POB.51., Hungary}
\author{T. Rauscher}
\affiliation{Department of Physics, University of Basel, CH-4056 Basel, Switzerland}

\date{\today}

\begin{abstract}
Captures of $\alpha$ particles on the proton-richest Barium isotope, $^{130}$Ba, have been studied in order to provide cross section data for the modelling of the astrophysical $\gamma$ process. The cross sections of the $^{130}$Ba($\alpha$,$\gamma$)$^{134}$Ce and $^{130}$Ba($\alpha$,n)$^{133}$Ce reactions have been measured with the activation technique in the center-of mass energy range between 11.6 and 16\,MeV, close above the astrophysically relevant energies. As a side result, the cross section of the $^{132}$Ba($\alpha$,n)$^{135}$Ce reaction has also been measured. The results are compared with the prediction of statistical model calculations, using different input parameters such as $\alpha$+nucleus optical potentials. It is found that the ($\alpha$,n) data can be reproduced employing the standard $\alpha$+nucleus optical potential widely used in astrophysical applications. Assuming its validity also in the astrophysically relevant energy window, we present new stellar reaction rates for $^{130}$Ba($\alpha$,$\gamma$)$^{134}$Ce and $^{132}$Ba($\alpha$,$\gamma$)$^{136}$Ce and their inverse reactions calculated with the SMARAGD statistical model code. The highly increased $^{136}$Ce($\gamma$,$\alpha$)$^{132}$Ba rate implies that the $p$ nucleus $^{130}$Ba cannot directly receive contributions from the Ce isotopic chain. Further measurements are required to better constrain this result.
\end{abstract}


\maketitle
\section{Introduction}
\label{sec:intro}

\begin{figure*}
\includegraphics[angle=270,width=0.8\textwidth]{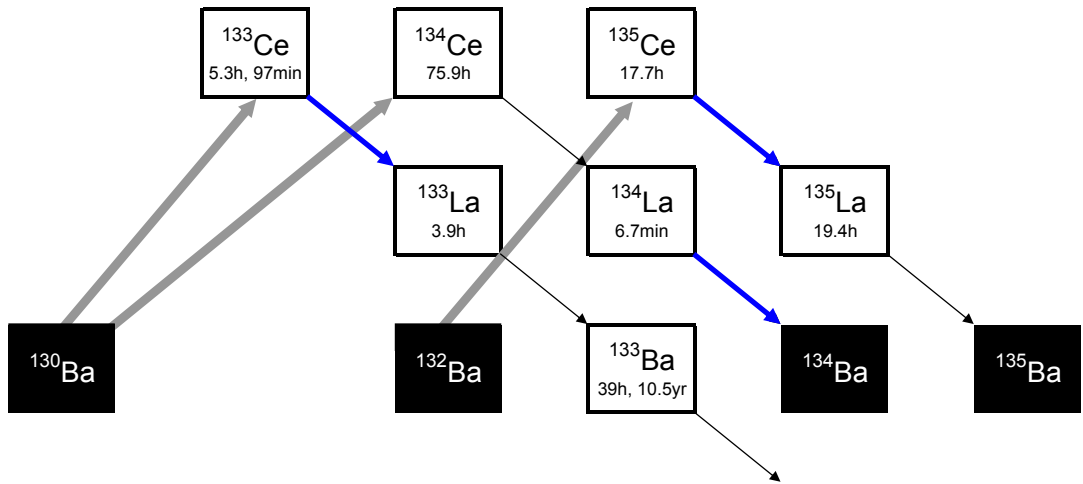}%
\caption{\label{fig:reactions} (color online) The investigated reactions and the decay of the reaction products. The three studied reactions are indicated by the thick grey arrows while the decays which have been used for the analysis are shown by the blue arrows. All the decay data are taken from refs. \cite{NDS133,NDS134,NDS135} with the exception of the $^{133}$Ce$^m$ half-life, where the newly determined value is used \cite{far11}.}
\end{figure*}

The synthesis of chemical elements heavier than iron is a complex problem in nuclear astrophysics \cite{jos11,thi11}. The origin of about 99\,\% of these heavy elements observed in the Solar System can be rather well explained by the classical slow \cite{kap06} and rapid \cite{arn07} neutron capture processes although details of these processes (especially the $r$ process) are still not completely known. The missing 1\,\% comprises those heavy, proton rich isotopes which cannot be created in neutron capture reactions. These are the so-called $p$ isotopes and in general their hypothetical production mechanism is referred to as the astrophysical $p$ process \cite{woo78}.

Since no single process has been found so far which could explain the observed abundances of all the $p$ isotopes, the $p$ process may actually involve several sub-processes, each of which may have its contribution to the production of at least a sub-group of $p$ isotopes. The possibilities include, e.g., the $rp$ process \cite{sch98}, the $\nu$ process \cite{woo90}, and the recently studied $\nu p$ process \cite{fro06}. The most important process is perhaps the so-called $\gamma$ process, which -- as its name suggests -- proceeds through $\gamma$-induced reactions \cite{woo78,ray95,arn03}.

If heavy seed isotopes produced earlier by the $s$ or $r$ processes are available in a hot stellar environment, then the high energy thermal photons may induce ($\gamma$,n) reactions driving the material towards the proton-rich isotopes. For more and more neutron deficient nuclei, also charged particles can be emitted. Then the ($\gamma$,p) and ($\gamma,\alpha$) reactions compete with the ($\gamma$,n) reactions and influence the reaction path. The required high temperatures (of the order of a few GK) are encountered in evolved massive stars either during the supernova explosion or shortly before that. Therefore, the $\gamma$ process is thought to take place in core-collapse supernova explosions, although other possible sites have also been suggested \cite{arn03}.

When the whole mass region of $p$ isotopes is considered, the modeling of the $\gamma$ process requires a huge reaction network of tens of thousands of reactions. The stellar rates of these reactions are required input to the network to calculate the $p$ isotope abundances. So far, no $\gamma$ process calculation in massive stars reproduced well the observed abundances of the $p$ isotopes simultaneously over the whole mass region. Since the reaction rates are typically taken from largely untested theoretical cross sections, the failure of the models may at least in part be attributed to incorrect reaction rates. Therefore, the experimental verification of the calculated cross sections is of high importance to provide better input for the network calculations and increase their reliability.

In order to reduce the complications caused by the thermal population of target states in the astrophysical plasma, it is preferable to study experimentally the inverse capture reactions instead of the $\gamma$-induced reactions \cite{rau10,xfactor}. Comparison of the available experimental data with the theoretical cross sections (which are typically obtained from Hauser-Feshbach type statistical model calculations) shows that for (p,$\gamma$) and (n,$\gamma$) reactions the models are able to reproduce the measured cross sections within about a factor of two. This can be considered as the general precision of the statistical models for these reactions, some fine-tuning of the parameters, however, being still possible \cite{kis07}. For ($\alpha,\gamma$) reactions, on the other hand, much larger deviations can be found. Moreover, there are only a handful of isotopes for which ($\alpha,\gamma$) cross section data are available in the mass range relevant for the $\gamma$ process. This severely limits the possibility of testing the model calculations (for a recent list of measured reaction see \cite{gyu10}). Therefore, measurements of $\alpha$ capture reaction cross sections in the relevant mass and energy range may help increase the reliability of statistical model calculations and lead to a better modeling of the $\gamma$ process.

Not all the reactions in a $\gamma$ process network have equal importance for the resulting $p$ isotope abundances. By carrying out a $\gamma$ process network calculation and varying the involved reaction rates, one can pinpoint those reactions to which the calculations show the highest sensitivity. Such a study has been carried out recently and a list of isotopes has been suggested on which proton or $\alpha$ capture cross sections should be measured \cite{rau06}. From the 15 naturally occurring isotopes for which high priority has been assigned to ($\alpha,\gamma$) measurements, the $p$ isotope of the element barium, $^{130}$Ba, has been selected for the present study. No experimental data is available yet in the literature for $\alpha$-induced reactions on $^{130}$Ba.

It can be shown (e.g., \cite{tomreview} and also the discussion in Sec.\ \ref{sec:comparison} below) that the ($\alpha,\gamma$) cross sections at astrophysical energies almost exclusively depend only on the $\alpha$ widths. Therefore they show a strong sensitivity to the $\alpha$+nucleus optical potential in statistical model calculations. Global optical potentials designed for a wide range of isotopes are typically used in astrophysical applications. Several global potentials are available in the literature and different potentials lead to largely different cross sections. The comparison with experimental data can aid the selection of the best optical potential for the description of low energy $\alpha$ capture reactions in the heavy element range and thus the collection of more experimental data at low energy is useful also for a better understanding of the optical potential.

In the following, we present further details of the investigated reactions in Sec.\ \ref{sec:details}. The experimental procedure is thoroughly explained in Sec.\ \ref{sec:exp} and the resulting cross sections are presented in Sec.\ \ref{sec:results}. The impact of the data on predictions of astrophysical reaction rates for the reactions $^{130}$Ba($\alpha$,$\gamma$)$^{134}$Ce and $^{134}$Ce($\gamma$,$\alpha$)$^{130}$Ba is discussed in Sec.\ \ref{sec:comparison}. Tables of the recommended reaction rates and their fits are presented in Sec.\ \ref{sec:conclusion}, along with some concluding remarks.

\section{The investigated reactions}
\label{sec:details}

The aim of the present work is the measurement of the $^{130}$Ba($\alpha,\gamma$)$^{134}$Ce reaction cross section. The relevant energy range (the Gamow window) for this reaction at a typical $\gamma$ process temperature of 2\,GK is between 5.3 and 8.1\,MeV \cite{rau10b}. In this energy range the ($\alpha,\gamma$) cross section is very low, the calculations with the NON-SMOKER code yield values between 10$^{-15}$ and 10$^{-8}$\,barns. This is unfortunately not measurable in the laboratory, thus the cross section must be measured at higher energies. Determined by the cross section values, the background conditions, and other experimental aspects (see below), the cross sections have been measured at 11 energies in the center-of-mass energy range between 11.6 and 16 MeV.

Alpha capture on $^{130}$Ba leads to $^{134}$Ce which is radioactive and decays to $^{134}$La with a half-life of 75.9\,h. Subsequently, $^{134}$La decays  to $^{134}$Ba with a half-life of 6.67\,m. This sequence allows the cross section to be measured with the activation technique which already has been used successfully for measurements relevant for the $\gamma$ process \cite{gyu10}. The decay of $^{134}$Ce is followed by the emission of $\gamma$ rays with a very low relative intensity of well below 1\,\%. On the other hand, $\gamma$ radiation of 5\,\% relative intensity accompanies the decay of $^{134}$La. The detection of this $\gamma$ radiation has been used to determine the cross section.

Besides radiative capture, several other reaction channels are open on $^{130}$Ba in the studied energy range. Only the ($\alpha$,n) and ($\alpha$,p) reactions have, however, non-negligible (above 1\,$\mu$b) cross sections. The cross section of $^{130}$Ba($\alpha$,n)$^{133}$Ce can again be measured with activation since both the ground state and the isomeric state of $^{133}$Ce decay with a suitable half-life, and the decays are followed by high intensity $\gamma$ radiation. This measurement has also been carried out and owing to the different decay patterns of the ground and isomeric states of $^{133}$Ce, partial cross section leading to these states could be determined separately.

The ($\alpha$,p) reaction can in principle also be measured with activation, but the product isotope, $^{133}$La, has a continuous feeding from the decay of the ($\alpha$,n) reaction product, $^{133}$Ce. Since the ($\alpha$,n) cross section is typically two orders of magnitude higher than the ($\alpha$,p) one, the production of $^{133}$La is entirely dominated by the decay of $^{133}$Ce  and therefore the ($\alpha$,p) cross section could not be separated.

The enriched $^{130}$Ba targets (see Sec.\ \ref{sec:target}) contained substantial amount of the $^{132}$Ba isotope as well. This isotope is also listed in Ref.\ \cite{rau06} in the second priority group of the isotopes on which $\alpha$ capture cross section measurements are needed for the $\gamma$ process. Radiative $\alpha$ capture on $^{132}$Ba leads to the stable $^{136}$La, hence its cross section cannot be measured with activation. The ($\alpha$,n) reaction, however, leads to the radioactive $^{135}$Ce. The cross section measurement of the $^{132}$Ba($\alpha$,n)$^{135}$Ce reaction has therefore been also carried out in the present work.

\begin{table}
\caption{\label{tab:reactions}Some decay parameters of the isotopes produced by the studied reactions. Only those $\gamma$ radiations are listed which have been used for the analysis. The data are taken from \cite{NDS133,NDS134,NDS135,far11}.}
\begin{ruledtabular}
\begin{tabular}{ccccc}
Reaction & Produced & Half-life & E$_\gamma$/keV & relative  \\
         & isotope &             &               &  $\gamma$ intensity [\%] \\
\hline
$^{130}$Ba($\alpha,\gamma$) & $^{134}$Ce & 75.9\,h & \multicolumn{2}{c}{only weak gammas} \\
$^{130}$Ba($\alpha,\gamma$) & $^{134}$La & 6.67\,min & 604.7 & 5.04\,$\pm$\,0.20 \\
$^{130}$Ba($\alpha$,n) & $^{133}$Ce$^g$ & 97\,min & 79.6 & 15.9\,$\pm$\,2.3 \\
                                             & & & 97.3 &  45.5\,$\pm$\,6.9 \\
                                             & & & 557.3 &  11.4\,$\pm$\,2.3 \\
$^{130}$Ba($\alpha$,n) & $^{133}$Ce$^m$ & 5.326\,h & 58.4 & 19.3\,$\pm$\,0.4 \\
                                             & & & 130.8 &  18.0\,$\pm$\,0.4 \\
                                             & & & 346.4 &  4.17\,$\pm$\,0.08 \\
                                             & & & 477.2 &  39.3\,$\pm$\,0.3 \\
                                             & & & 689.5 &  4.13\,$\pm$\,0.12 \\
                                             & & & 784.5 &  9.67\,$\pm$\,0.25 \\
$^{132}$Ba($\alpha$,n) & $^{135}$Ce & 17.7\,h & 265.6 & 41.8\,$\pm$\,1.4 \\
                                             & & & 300.1 &  23.5\,$\pm$\,0.5 \\
\end{tabular}
\end{ruledtabular}
\end{table}

The studied reactions and the decay of the reaction products are shown schematically in Fig.\ \ref{fig:reactions}. The relevant decay parameters of the produced isotopes used for the analysis are summarized in Table \ref{tab:reactions}. All decay parameters are taken from Refs.\ \cite{NDS133,NDS134,NDS135} with the exception of the half-life of $^{133}$Ce$^m$. The literature value of the $^{133}$Ce$^m$ half-life has an unusually high uncertainty and the analysis of our data indicated that the half-life is strongly underestimated. We have therefore carried out a dedicated half-life measurement of this isomeric state resulting in a much more precise value. Details and the result of this work has already been published elsewhere \cite{far11}.

\section{Experimental procedure}
\label{sec:exp}

\subsection{Target properties}
\label{sec:target}

As it is typical for the $p$ isotopes, the natural abundance of $^{130}$Ba is very low, only 0.1\,\%. The use of enriched targets was therefore necessary. The $^{130}$Ba content of the enriched material was 11.8\,$\pm$\,0.2\,\% and it contained also 0.7\,$\pm$\,0.1\,\% $^{132}$Ba. The enriched material was available in BaCO$_3$ form. Targets have been prepared by evaporating BaCO$_3$ onto thin Al foils. Since the stoichiometry of the compound may change during the evaporation process and the reliable determination of the number of $^{130}$Ba target atoms (i.e., the target thickness) is very important for the cross section measurements, three different methods have been used to measure the target thickness. The weight of the Al foils has been measured before and after the irradiation, the energy loss of $\alpha$ particles penetrating the targets has been measured and the target thickness has been also measured by Rutherford Backscattering Spectrometry. Details of the target thickness measurements have been published elsewhere \cite{gyu11}. Good agreement between the three methods have been found and a final value of 8\,\% could be assigned to the uncertainty of the number of target atoms. The number density of the $^{130}$Ba atoms was between 3\,$\cdot$\,10$^{16}$ and 2\,$\cdot$\,10$^{17}$ atoms/cm$^2$.

\subsection{Irradiations}
\label{sec:irrad}

The irradiations have been carried out at the cyclotron accelerator of ATOMKI. The targets have been put into the target chamber which had been used in several activation experiments before (see e.g. \cite{gyu06}).

The maximum beam current was restricted to about 1\,$\mu$A in order to avoid target deterioration. Nevertheless, the target stability has been continuously monitored by detecting the backscattered alphas with a particle detector built into the chamber.

The length of the irradiations varied between several hours and two days. Longer irradiations have been used at the lowest energies where the cross sections are low in order to increase the number of long lived $^{134}$Ce nuclei. The half-life of the $^{130}$Ba($\alpha$,n) reaction products are, however, short compared to the length of the irradiation. Therefore, the time dependence of the beam current during the irradiation must be known. For this purpose the beam current was integrated and the number of current integrator counts was recorded in multichannel scaling with a time basis of one minute.

\subsection{Detection of the induced $\gamma$ radiation}
\label{sec:detection}

\begin{figure}
\includegraphics[width=\columnwidth]{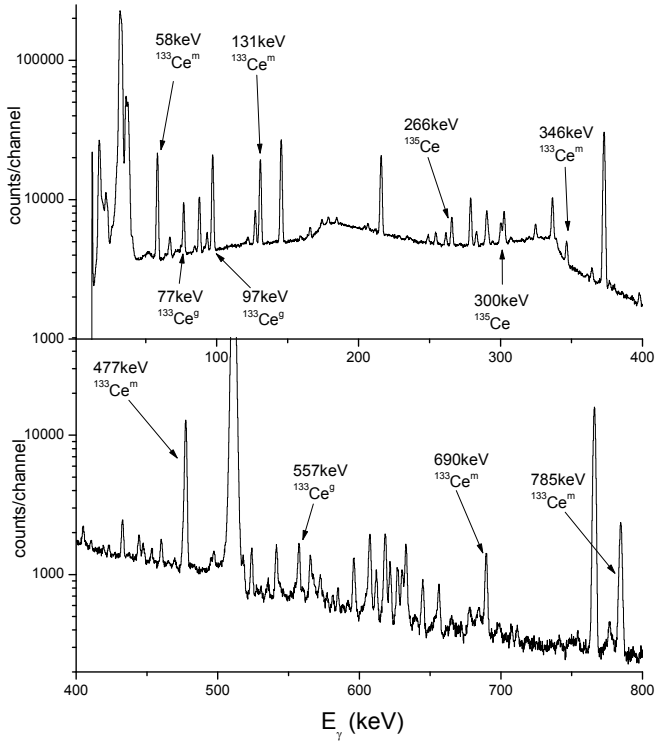}%
\caption{\label{fig:anspectrum} Typical $\gamma$ spectrum taken on a target irradiated by a 14\,MeV $\alpha$ beam shortly after the end of the irradiation. The peaks used for the cross section determination are indicated by arrows.}
\end{figure}

\begin{figure}
\includegraphics[angle=270,width=\columnwidth]{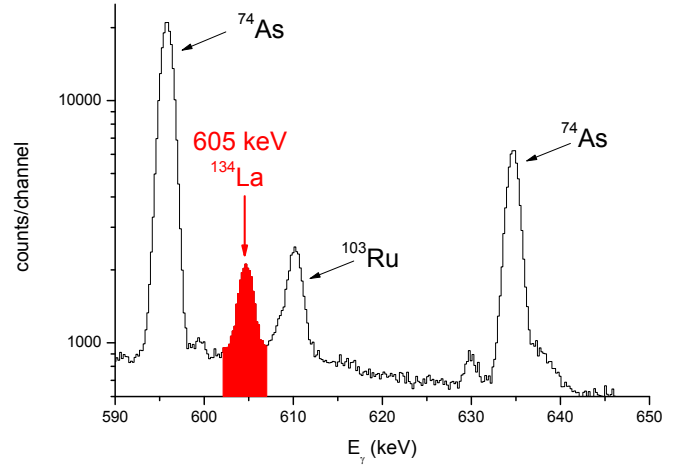}%
\caption{\label{fig:agspectrum} (color online) The relevant part of the $\gamma$ spectrum measured on a target irradiated by a 14\,MeV $\alpha$ beam. The red area and label indicate the peak from the decay of $^{134}$La used for the $^{130}$Ba($\alpha$,$\gamma$)$^{134}$Ce cross section measurement. Some other peaks from radioisotopes produced on target impurities are also indicated.}
\end{figure}

\begin{figure}
\includegraphics[angle=270,width=\columnwidth]{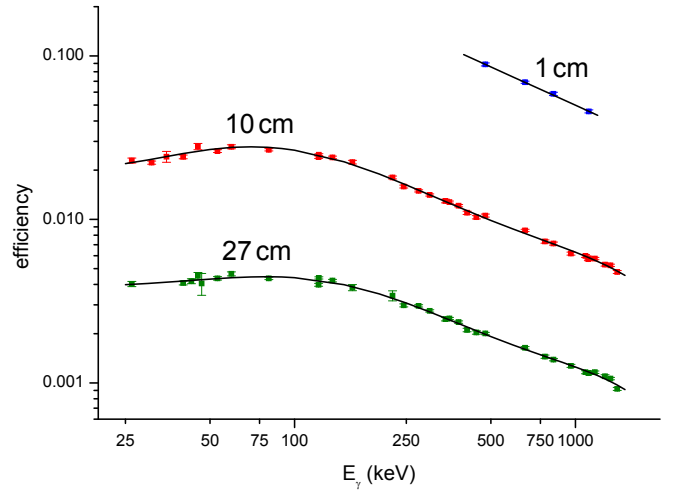}%
\caption{\label{fig:efficiency} (color online) The measured and fitted absolute efficiency of the HPGe detector. In the 1\,cm geometry $^{7}$Be, $^{54}$Mn, $^{65}$Zn, and $^{137}$Cs single line sources have been used in order to avoid true coincidence summing. In the 10\,cm geometry $^{22}$Na, $^{57}$Co, $^{60}$Co, $^{133}$Ba, $^{152}$Eu, and $^{241}$Am sources have been used additionally.}
\end{figure}

After the end of the irradiations the targets have been transported to the low-background counting setup of ATOMKI. This setup consists of a 100\,\% relative efficiency Canberra HPGe detector and a complete 4$\pi$ low background shielding \cite{far11b}.

Owing to the largely different half-lives of the reaction products of the different investigated reactions, two separate $\gamma$ countings have been carried out. The first counting started typically 30 minutes after the end of the irradiation. In this phase the decay of the $^{133}$Ce$^g$, $^{133}$Ce$^m$, and $^{135}$Ce isotopes could be measured.

Figure \ref{fig:anspectrum} shows a typical $\gamma$ spectrum taken in this first counting period on a target irradiated by a 14\,MeV $\alpha$ beam. The $\gamma$ lines used for the analysis, as listed in Table \ref{tab:reactions}, are indicated by arrows. Many other peaks are visible in the spectrum, they belong either to the decay of the studied isotopes but not used in the analysis, to other isotopes induced by reactions on heavier Ba isotopes, or to radioisotopes produced on target impurities. The spectra were stored regularly, every 10 minutes in the first five hours and every hour afterwards. The decay of the different isotopes was followed in this way and a half-life analysis guaranteed that no short- or long-lived isotopes gave contribution to the studied peaks.

After several days of cooling, the $\gamma$ spectra of the targets have been taken again in order to measure the $^{130}$Ba($\alpha$,$\gamma$)$^{134}$Ce cross section through the decay of $^{134}$La. The relevant part of the spectrum measured on a target irradiated at 14\,MeV can be seen in Fig. \ref{fig:agspectrum}. The length of the counting was four days and it started approximately six days after the end of the irradiation. The peak of the $^{134}$La decay used for the $^{130}$Ba($\alpha$,$\gamma$)$^{134}$Ce cross section determination is indicated by the red area and label. Some other peaks are also indicated in the figure. They correspond to the decay of $^{74}$As and $^{103}$Ru, which have been produced most likely by the $^{71}$Ga($\alpha$,n)$^{74}$As and $^{100}$Mo($\alpha$,n)$^{103}$Ru reactions, respectively, induced on target impurities.

Three different counting geometries have been used in the course of the measurements. The samples have been put either at 1\,cm, 10\,cm or 27\,cm distance from the end cap of the detector in its geometric axis. The absolute efficiency of the detector has been measured in all three counting geometries with several calibrated radioactive sources. The used sources were the following: $^{7}$Be, $^{22}$Na, $^{54}$Mn, $^{57}$Co, $^{60}$Co, $^{65}$Zn, $^{133}$Ba, $^{137}$Cs, $^{152}$Eu, $^{241}$Am. Owing to the large true coincidence summing effect at close geometries, only single line calibration sources ($^{7}$Be, $^{54}$Mn, $^{65}$Zn, $^{137}$Cs) have been used at the 1\,cm geometry.

Figure \ref{fig:efficiency} shows the measured efficiencies. At 10 and 27\,cm, the measured points have been fitted with a sixth order logarithmic polynomial. At 1\,cm the measured point fall into the energy region where the efficiency curve can well be described with a log-log straight line. The fitted curves can also be seen in the figure.

Since the decay of the $^{130}$Ba($\alpha$,n)$^{133}$Ce reaction products involves a high number of mainly low-energy $\gamma$ transitions, the first countings have been carried out in the 10\,cm geometry in order to avoid large true coincidence summing effects. Nevertheless, even at this distance the summing effect cannot be neglected and must be determined. This was carried out by measuring the spectrum of a high activity $^{133}$Ce source (produced in a high-energy $\alpha$ irradiation of a $^{130}$Ba target) both in the 27 and 10\,cm geometries. At 27\,cm the summing effect is well below 1\,\% and therefore neglected. By taking into account the time elapsed betwen the two countings, a summing correction factor for all the studied transitions was calculated. This corrections factor was always below 10\,\%.

In order to maximize the count rate for the low intensity 605\,keV transition of the $^{134}$La decay, the second countings have been carried out in the 1\,cm geometry. In principle, the decay of $^{134}$La is free from true coincidence summing, since the 605\,keV $\gamma$ line represents the only strong transition. The nucleus $^{134}$La is, however, a positron emitter. Therefore, the true coincidence between the 511\,keV positron annihilation line and the 605\,keV line cannot be neglected. A similar procedure has thus been followed as in the case of the $^{130}$Ba($\alpha$,n)$^{133}$Ce measurements. Several strong $^{134}$La sources have been prepared by the $^{134}$Ba(p,n)$^{134}$La reaction. For this purpose, natural BaCO$_3$ and BaCl$_2$ targets have been irradiated by 8\,MeV protons at the cyclotron of ATOMKI. The $\gamma$ spectra at 10 and 1\,cm have been measured and a correction factor of 30\,\%\,$\pm$\,1.3\,\% has been obtained.

\section{Cross section results}
\label{sec:results}

\begin{table}
\caption{\label{tab:130ag}Measured cross section of the $^{130}$Ba($\alpha$,$\gamma$)$^{134}$Ce reaction}
\begin{tabular*}{\columnwidth}{c@{\extracolsep{\fill}}r@{\extracolsep{1mm}}c@{\extracolsep{1mm}}l@{\extracolsep{\fill}}r@{\extracolsep{1mm}}c@{\extracolsep{1mm}}l}
\hline
E$_{\rm{beam}}$ & \multicolumn{3}{c}{E$_{\rm{c.m.}}$} & \multicolumn{3}{c}{cross section} \\
MeV & \multicolumn{3}{c}{MeV} & \multicolumn{3}{c}{$\mu$barn} \\
\hline
12.0    &       11.613  &       $\pm$   &       0.045   &       65.5    &       $\pm$   &       9.8     \\
12.5    &       12.133  &       $\pm$   &       0.052   &       118.8   &       $\pm$   &       12.7    \\
13.0    &       12.596  &       $\pm$   &       0.041   &       328     &       $\pm$   &       48      \\
13.5    &       13.080  &       $\pm$   &       0.042   &       410     &       $\pm$   &       48      \\
14.0    &       13.559  &       $\pm$   &       0.046   &       940     &       $\pm$   &       67      \\
14.5    &       14.024  &       $\pm$   &       0.060   &       1453    &       $\pm$   &       108     \\
15.0    &       14.509  &       $\pm$   &       0.061   &       2051    &       $\pm$   &       187     \\
15.5    &       14.995  &       $\pm$   &       0.061   &       2051    &       $\pm$   &       146     \\
16.0    &       15.509  &       $\pm$   &       0.048   &       1937    &       $\pm$   &       142     \\
16.5    &       15.998  &       $\pm$   &       0.049   &       1567    &       $\pm$   &       114     \\
\hline
\end{tabular*}
\end{table}

\begin{table*}
\caption{\label{tab:130an}Measured cross section of the $^{130}$Ba($\alpha$,n)$^{133}$Ce reaction}
\begin{tabular}{c@{\extracolsep{1cm}}r@{\extracolsep{1mm}}c@{\extracolsep{1mm}}l@{\extracolsep{1cm}}r@{\extracolsep{1mm}}c@{\extracolsep{1mm}}l@{\extracolsep{1cm}}r@{\extracolsep{1mm}}c@{\extracolsep{1mm}}l@{\extracolsep{1cm}}r@{\extracolsep{1mm}}c@{\extracolsep{1mm}}l}
\hline
 & & & & \multicolumn{3}{c}{ground state} & \multicolumn{3}{c}{isomeric state} & \multicolumn{3}{c}{total} \\
E$_{\rm{beam}}$ & \multicolumn{3}{c}{E$_{\rm{c.m.}}$} & \multicolumn{3}{c}{cross section} & \multicolumn{3}{c}{cross section} & \multicolumn{3}{c}{cross section}\\
MeV & \multicolumn{3}{c}{MeV} & \multicolumn{3}{c}{$\mu$barn} & \multicolumn{3}{c}{$\mu$barn} & \multicolumn{3}{c}{$\mu$barn}\\
\hline
12.5    &       12.051  &       $\pm$   &       0.083   &       411     &       $\pm$   &       82      &       249     &       $\pm$   &       36      &       660     &       $\pm$   &       116     \\
13.0    &       12.596  &       $\pm$   &       0.041   &       983     &       $\pm$   &       145     &       611     &       $\pm$   &       68      &       1594    &       $\pm$   &       206     \\
13.5    &       13.080  &       $\pm$   &       0.042   &       2104    &       $\pm$   &       310     &       1620    &       $\pm$   &       181     &       3724    &       $\pm$   &       469     \\
14.0    &       13.559  &       $\pm$   &       0.046   &       6960    &       $\pm$   &       1013    &       5519    &       $\pm$   &       613     &       12479   &       $\pm$   &       1549    \\
14.5    &       14.024  &       $\pm$   &       0.060   &       14502   &       $\pm$   &       2111    &       10905   &       $\pm$   &       1212    &       25408   &       $\pm$   &       3164    \\
15.0    &       14.509  &       $\pm$   &       0.061   &       23762   &       $\pm$   &       3459    &       22563   &       $\pm$   &       2506    &       42651   &       $\pm$   &       5211    \\
15.5    &       14.995  &       $\pm$   &       0.061   &       34611   &       $\pm$   &       5041    &       36151   &       $\pm$   &       4016    &       70762   &       $\pm$   &       8584    \\
16.0    &       15.509  &       $\pm$   &       0.048   &       37830   &       $\pm$   &       5511    &       43548   &       $\pm$   &       4837    &       81378   &       $\pm$   &       9803    \\
16.5    &       15.998  &       $\pm$   &       0.049   &       37941   &       $\pm$   &       5524    &       44203   &       $\pm$   &       4911    &       82144   &       $\pm$   &       9892    \\
\hline
\end{tabular}
\end{table*}

\begin{table}
\caption{\label{tab:132an}Measured cross section of the $^{132}$Ba($\alpha$,n)$^{135}$Ce reaction}
\begin{tabular*}{\columnwidth}{c@{\extracolsep{\fill}}r@{\extracolsep{1mm}}c@{\extracolsep{1mm}}l@{\extracolsep{\fill}}r@{\extracolsep{1mm}}c@{\extracolsep{1mm}}l}
\hline
E$_{\rm{beam}}$ & \multicolumn{3}{c}{E$_{\rm{c.m.}}$} & \multicolumn{3}{c}{cross section} \\
MeV & \multicolumn{3}{c}{MeV} & \multicolumn{3}{c}{$\mu$barn} \\
\hline
12.5    &       12.051  &       $\pm$   &       0.083   &       815     &       $\pm$   &       159     \\
13.0    &       12.596  &       $\pm$   &       0.041   &       1800    &       $\pm$   &       435     \\
13.5    &       13.080  &       $\pm$   &       0.042   &       4520    &       $\pm$   &       923     \\
14.0    &       13.559  &       $\pm$   &       0.046   &       11900   &       $\pm$   &       2194    \\
14.5    &       14.024  &       $\pm$   &       0.060   &       21500   &       $\pm$   &       3959    \\
15.0    &       14.509  &       $\pm$   &       0.061   &       55100   &       $\pm$   &       10115   \\
15.5    &       14.995  &       $\pm$   &       0.061   &       82000   &       $\pm$   &       15053   \\
16.0    &       15.509  &       $\pm$   &       0.048   &       95400   &       $\pm$   &       17513   \\
16.5    &       15.998  &       $\pm$   &       0.049   &       111000  &       $\pm$   &       20180   \\
\hline
\end{tabular*}
\end{table}

Tables \ref{tab:130ag}, \ref{tab:130an}, and \ref{tab:132an} show the measured cross sections of the three investigated reactions. In the case of the $^{130}$Ba($\alpha$,n)$^{133}$Ce reactions, partial cross sections for the population of the ground and isomeric states have been determined separately. These are listed in Table \ref{tab:130an} as well as the total cross section.

The effective center-of-mass energies indicated in the second columns have been calculated taking into account the beam energy loss in the target. From the known target thickness the energy loss has been calculated with the SRIM code \cite{SRIM}. Depending on the target and the beam energy the energy loss was between about 20 and 150\,keV. Since the cross section is a smoothly varying function of the energy in the studied energy range and the cross section does not change much within the thicknesses of the target, the effective energy has been chosen to correspond to the middle of the target. For the uncertainty of the effective energy the quadratic sum of half of the target thickness and the uncertainty of the beam energy has been taken, the latter one being 0.3\,\%.

In the case of those reactions where more than one $\gamma$ transition have been analyzed the results were always in good agreement. The error weighted mean has been calculated and adopted as the final cross section result. The uncertainty of the cross section values has been calculated using error propagation based on the following components: counting statistics ($\leq$\,15\,\%), decay parameters ($\leq$\,15\,\%), number of target atoms (8\,\%) \cite{gyu11}, detector efficiency (5\,\%), summing correction factor ($\leq$\,5\,\%), beam current integration (3\,\%). The uncertainty of the $^{132}$Ba($\alpha$,n)$^{135}$Ce cross section is somewhat higher owing to the large uncertainty (14\,\%) of the $^{132}$Ba isotopic abundance in the enriched $^{130}$Ba material.

\section{Impact on predictions of astrophysical reaction rates}
\label{sec:comparison}

\begin{figure}
\includegraphics[angle=270,width=\columnwidth]{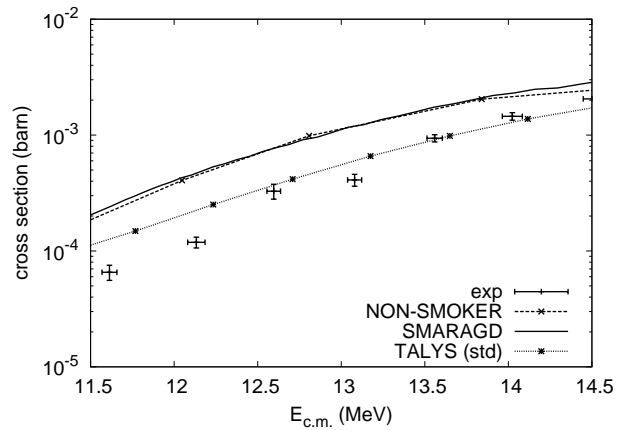}%
\caption{\label{fig:xs_ag} Comparison of experimental cross sections for $^{130}$Ba($\alpha$,$\gamma$)$^{134}$Ce and theoretical predictions with the codes NON-SMOKER \cite{nonsmoker,adndt}, SMARAGD \cite{SMARAGD,tomreview}, and TALYS \cite{TALYS} (using their default settings).}
\end{figure}

\begin{figure}
\includegraphics[angle=270,width=\columnwidth]{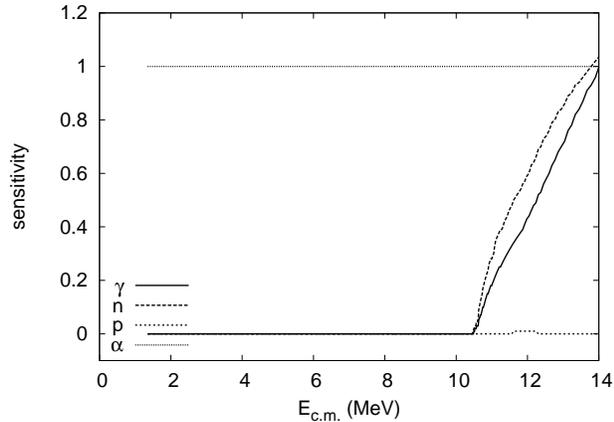}%
\caption{\label{fig:xs_sensi} Sensitivity of the cross sections for $^{130}$Ba($\alpha$,$\gamma$)$^{134}$Ce on variation of the neutron-, proton-, $\alpha$-, and $\gamma$ width.}
\end{figure}

\begin{figure}
\includegraphics[angle=270,width=\columnwidth]{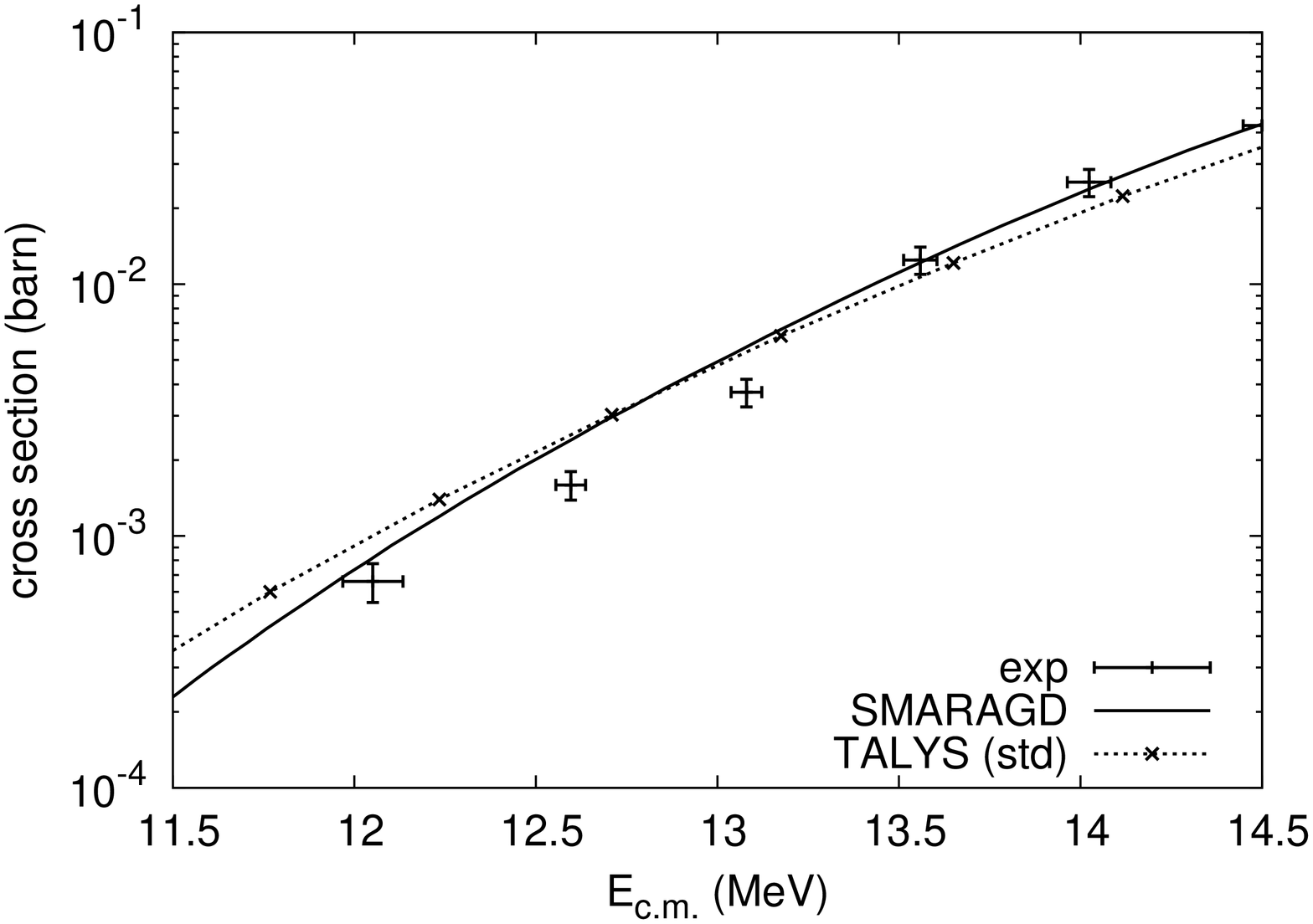}%
\caption{\label{fig:xs_an} Comparison of experimental cross sections for $^{130}$Ba($\alpha$,n)$^{133}$Ce and theoretical predictions with the codes SMARAGD \cite{SMARAGD,tomreview} and TALYS \cite{TALYS}
(using their default settings).}
\end{figure}

\begin{figure}
\includegraphics[angle=270,width=\columnwidth]{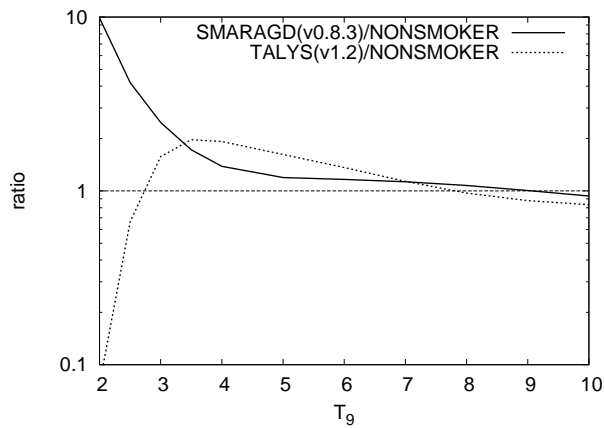}%
\caption{\label{fig:ratecomp} Ratios of astrophysical reaction rates for $^{130}$Ba($\alpha$,$\gamma$)$^{134}$Ce obtained with the SMARAGD and TALYS codes to the standard NON-SMOKER rate \cite{adndt} as function of astrophysical temperature $T_9$ (given in GK).}
\end{figure}

The experimental cross sections for the reaction $^{130}$Ba($\alpha$,$\gamma$)$^{134}$Ce are compared to theoretical predictions with the default results from the Hauser-Feshbach statistical model codes NON-SMOKER \cite{nonsmoker,adndt}, SMARAGD (version 0.8.3s) \cite{SMARAGD,tomreview}, and TALYS (version 1.2) \cite{TALYS} in Fig.\ \ref{fig:xs_ag}. The NON-SMOKER and SMARAGD predictions are too high by factors of $2-3$ within the shown energy range. Due to additional reaction mechanisms, such as possible pre-equilibrium particle emission, the results of these codes cannot be compared to the data above 14 MeV for this reaction but this is not relevant for the astrophysical conclusions, as discussed below. The TALYS prediction seems to be closer to the data in the upper part of the shown energy range but its energy dependence is different from the one shown by the data, leading also to an overprediction of almost a factor of two at the lowest energy.

It is not possible to draw a straighforward conclusion on the prediction of astrophysical reaction rates from this comparison. A closer inspection of the sensitivity of the resulting cross sections on a variation of the calculated transmission coefficients (averaged widths) is necessary to understand the source of the discrepancies and its possible impact on the reaction rate.
This sensitivity is shown in Fig.\ \ref{fig:xs_sensi} for a factor two variation of the widths. A sensitivity of zero implies that the cross section does not change when varying the width, a factor of unity means that the cross section changes by the same factor as the variation factor used for the width \cite{tomreview}. In the measured energy range, the cross sections are sensitive to three widths, the neutron-, $\gamma$-, and $\alpha$ width, respectively.

From the $\alpha$ capture measurement alone it would not be possible to further constrain the origin of the deviation from the data because of the sensitivity to several widths. The combination with the ($\alpha$,n) data in the same energy range, however, allows to check the validity of the $\alpha$ width because the ($\alpha$,n) cross section is mostly sensitive to this width at these energies. The comparison of calculated to measured cross sections for the reaction $^{130}$Ba($\alpha$,n)$^{133}$Ce is shown in Fig.\ \ref{fig:xs_an}. As can be seen, there is good agreement with the SMARAGD prediction. 
This implies that the $\alpha$ transmission coefficient is predicted well and is not the reason for the overpredicted ($\alpha$,$\gamma$) cross section. The code TALYS also predicts the absolute value of the ($\alpha$,n) cross section well in this energy range although it is using a different optical $\alpha$+nucleus potential. The energy dependence, however, is different than the one of the data and thus it seems fortuitous that the absolute value is reproduced within the shown energy range. The deviation in energy dependence, especially towards lower energies, is less strong than the one found in the ($\alpha$,$\gamma$) case, implying that also the energy dependence of the $\gamma$- and/or neutron width may not be correctly described at low energy. A different $\alpha$ width energy dependence will lead to quite different cross sections, compared to the SMARAGD prediction, at lower -- astrophysically important -- energies.

Combining the conclusions from the ($\alpha$,$\gamma$) and ($\alpha$,n) data it is obvious that the difference between the SMARAGD and TALYS predictions for the $\alpha$ capture stems from differently predicted neutron- or/and $\gamma$ widths. With the given data it is not possible to further constrain the uncertainties. But are these widths actually astrophysically relevant?
The astrophysical energy window for $^{130}$Ba($\alpha$,$\gamma$)$^{134}$Ce is about $5-8$ MeV \cite{rau10b}. Fig.\ \ref{fig:xs_sensi} shows that the $\alpha$ width is the only determining factor in this energy range, as it is by far smaller than the $\gamma$ width. Therefore the uncertainty in the prediction of the $\alpha$ width will dominate the resulting uncertainty in the prediction of the astrophysical reaction rate.

A comparison of the astrophysical rates obtained with the various codes is presented in Fig.\ \ref{fig:ratecomp}. Since the majority of astrophysical simulations make use of reaction rates from \cite{adndt} which were obtained with the NON-SMOKER code, the ratio of the more recent predictions to this standard is plotted. The predictions differ by not more than 20\% at the highest temperatures but they run apart at lower temperatures \cite{talysrates}. The relevant temperature region is $2-3$ GK, at the lower end of the shown range. It is noteworthy that these results were obtained with the standard settings of the codes, not attempting to fit any further energy dependence of the optical $\alpha$+nucleus potential to the data. It has to be realized that even the reproduction of the energy dependence of the averaged $\alpha$ width in the measured energy range -- as provided by the potential by \cite{McF} used with the SMARAGD and NON-SMOKER codes here -- does not guarantee that the energy dependence is similarly well described at much lower energies.
This underlines the fact that it is necessary to measure within the astrophysically relevant energy window in order to constrain
the reaction rates.

\begin{figure}
\includegraphics[angle=270,width=\columnwidth]{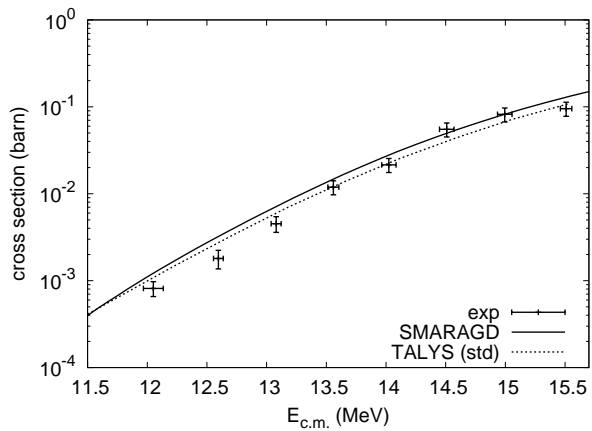}%
\caption{\label{fig:xs_ba132an} Comparison of experimental cross sections for $^{132}$Ba($\alpha$,n)$^{135}$Ce and theoretical predictions with the codes SMARAGD \cite{SMARAGD,tomreview} and TALYS \cite{TALYS} (using their default settings).}
\end{figure}

\begin{figure}
\includegraphics[angle=270,width=\columnwidth]{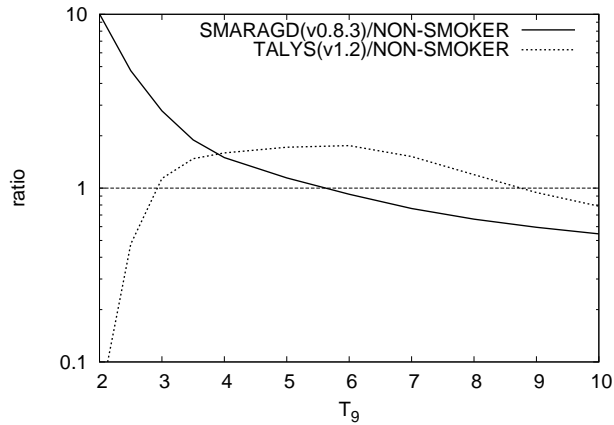}%
\caption{\label{fig:ratecomp132} Ratios of astrophysical reaction rates for $^{132}$Ba($\alpha$,$\gamma$)$^{136}$Ce obtained with the SMARAGD and TALYS codes to the standard NON-SMOKER rate \cite{adndt} as function of astrophysical temperature $T_9$ (given in GK).}
\end{figure}

A very similar picture arises when comparing the measured $^{132}$Ba($\alpha$,n)$^{135}$Ce cross sections to predictions, as shown in Fig.\ \ref{fig:xs_ba132an}. The theoretical calculations reproduce the data well (with similar deviations from the measured energy dependence as in the $^{130}$Ba case) which leads to the conclusion that the $\alpha$ widths are described well in the measured energy range. Also the comparison of the actual stellar reaction rates for $^{132}$Ba($\alpha$,$\gamma$)$^{136}$Ce obtained with the different codes in Fig.\ \ref{fig:ratecomp132} looks similar to the previous case: the newer predictions are lower at high temperature than the NON-SMOKER rates but show a different temperature dependence. While the SMARAGD rate is considerably higher in the relevant temperature range, the TALYS rate is comparable or lower than the standard rate.

\section{Conclusions}
\label{sec:conclusion}

The cross sections for the reactions $^{130}$Ba($\alpha$,$\gamma$)$^{134}$Ce, $^{130}$Ba($\alpha$,n)$^{133}$Ce, and $^{132}$Ba($\alpha$,n)$^{135}$Ce have been measured closely above the ($\alpha$,n) threshold with the activation technique. The combination of ($\alpha$,$\gamma$) and ($\alpha$,n) data allowed to test the predicted $\alpha$ widths (and thus the optical $\alpha$+nucleus potential) in the measured energy range. Good reproduction of the $^{130}$Ba($\alpha$,n)$^{133}$Ce and $^{132}$Ba($\alpha$,n)$^{135}$Ce cross sections was found when using the potential by \cite{McF}, which is a standard potential often used in astrophysical applications.

Assuming the validity of the potential by \cite{McF} also at low energies, we present Table \ref{tab:reactivity} with the stellar reactivities obtained with the SMARAGD code, version 0.8.3s, for the $^{130}$Ba($\alpha$,$\gamma$)$^{134}$Ce reaction and its inverse. The parameters for a fit of the reactivities in REACLIB format are given in Table \ref{tab:ratefit}. Stellar reactivities include transitions from thermally populated states of the target nuclei.  Even if it were possible to measure in the astrophysically relevant energy range, only the transitions from the ground state of the target would be determined. The ground state contribution $X$ to the stellar rate as introduced in Ref.\ \cite{xfactor} is also given, showing by how much the uncertainty may be reduced by a measurement on $^{130}$Ba or $^{134}$Ce, respectively, in the ground state. Comparing $X_{\alpha \gamma}$ for the capture reaction with $X_{\gamma \alpha}$ for the reverse ($\gamma$,$\alpha$) reaction also nicely shows how much smaller the measured contribution to the stellar rate would be when using a photodisintegration experiment instead of a capture one.

\begin{table*}
\caption{\label{tab:reactivity}Stellar reactivities $R$ in cm$^3$s$^{-1}$mole$^{-1}$ and ground state contributions $X$ to the stellar rate as function of temperature $T$. Both quantities are given for the reactions $^{130}$Ba($\alpha$,$\gamma$)$^{134}$Ce and $^{134}$Ce($\gamma$,$\alpha$)$^{130}$Ba.}
\begin{ruledtabular}
\begin{tabular}{ddddd}
\multicolumn{1}{c}{$T$ (GK)}& \multicolumn{1}{c}{$R_{\alpha \gamma}$}& \multicolumn{1}{c}{$X_{\alpha \gamma}$}&\multicolumn{1}{c}{$R_{\gamma \alpha }$}& \multicolumn{1}{c}{$X_{\gamma \alpha }$}\\
\hline
  0.15 &  1.173 \times 10^{-84} & 1.0 &   5.4699\times 10^{-75}   &     0.1882 \\
  0.20 &  1.137 \times 10^{-72} & 1.0 &   8.0374\times 10^{-63}   &     0.1787 \\
  0.30 &  1.363 \times 10^{-58} & 1.0 &   1.7428\times 10^{-48}   &     0.1501 \\
  0.40 &  7.683\times 10^{-50} & 0.9998 & 1.5008\times 10^{-39}  &  0.1318 \\
  0.50 &  1.094\times 10^{-43} & 0.9988 & 2.9757\times 10^{-33}  &  0.1197 \\
  0.60 &  4.804\times 10^{-39} & 0.9950 & 1.7158\times 10^{-28}  &  0.1105 \\
  0.70 &  2.235\times 10^{-35} & 0.9868 & 1.0086\times 10^{-24}  &  0.1032 \\
  0.80 &  2.204\times 10^{-32} & 0.9727 & 1.2250\times 10^{-21}  &  0.09701 \\
  0.90 &  7.044\times 10^{-30} & 0.9522 & 4.7184\times 10^{-19}  &  0.09150 \\
  1.00 &  9.621\times 10^{-28} & 0.9259 & 7.6701\times 10^{-17}  &  0.08637 \\
  1.50 &  2.119\times 10^{-20} & 0.7439 & 3.4747\times 10^{-9}  &  0.05929 \\
  2.00 &  5.827\times 10^{-16} & 0.5558 & 1.6807\times 10^{-4}  &  0.03129 \\
  2.50 &  6.798\times 10^{-13} & 0.4032 & 3.0934\times 10^{-1}  &  0.01398 \\
  3.00 &  1.378\times 10^{-10} & 0.2911 & 9.1856\times 10^{1}  &  0.005829 \\
  3.50 &  9.264\times 10^{-9} & 0.2126 &  8.6161\times 10^{3}  &   0.002428 \\
  4.00 &  2.783\times 10^{-7} & 0.1594 &  3.4934\times 10^{5}  &   0.001085 \\
  4.50 &  4.397\times 10^{-6} & 0.1230 &  7.1454\times 10^{6}  &   5.396 \times 10^{-4} \\
  5.00 &  4.096\times 10^{-5} & 0.09718 & 7.8862\times 10^{7}  &  2.971\times 10^{-4} \\
  6.00 &  1.010\times 10^{-3} & 0.06354 & 1.8376\times 10^{9}  &  1.096\times 10^{-4} \\
  7.00 &  7.134\times 10^{-3} & 0.04276 & 1.0434\times 10^{10}  &  4.365\times 10^{-5} \\
  8.00 &  2.278\times 10^{-2} & 0.02889 & 3.1260\times 10^{10}  &  1.573\times 10^{-5} \\
  9.00 &  4.550\times 10^{-2} & 0.01923 & 6.5428\times 10^{10}  &  4.662\times 10^{-6} \\
 10.00 &  6.855\times 10^{-2} & 0.01250 & 1.0774\times 10^{11}  &  1.168\times 10^{-6}
\end{tabular}
\end{ruledtabular}
\end{table*}

\begin{table*}
\caption{\label{tab:ratefit}Fit parameters for the REACLIB fit format \cite{adndt} for $^{130}$Ba($\alpha$,$\gamma$)$^{134}$Ce and its inverse. Note that the value for $^{134}$Ce($\gamma$,$\alpha$)$^{130}$Ba obtained in this parameterization has to be multiplied by the ratio of the temperature-dependent partition functions $G(^{130}\mathrm{Ba})/G(^{134}\mathrm{Ce})$ \cite{adndt} to get the stellar reactivity.}
\begin{ruledtabular}
\begin{tabular}{lddddddd}
&\multicolumn{1}{c}{$a_0$}&\multicolumn{1}{c}{$a_1$}&\multicolumn{1}{c}{$a_2$}&\multicolumn{1}{c}{$a_3$}&\multicolumn{1}{c}{$a_4$}&\multicolumn{1}{c}{$a_5$}&\multicolumn{1}{c}{$a_6$}\\
\hline
($\alpha$,$\gamma$)&-842.1489&-75.13541&-1326.36&2286.821&-112.6802&5.078861&-1117.492\\
($\gamma$,$\alpha$)&-817.10229989&-75.12610319&-1326.36&2286.821&-112.6802&5.078861&-1115.992
\end{tabular}
\end{ruledtabular}
\end{table*}

The reaction $^{134}$Ce($\gamma$,$\alpha$)$^{130}$Ba received special mention in \cite{rau06} because its rate is close (within a factor of 10) to the $^{134}$Ce($\gamma$,n)$^{133}$Ce rate at plasma temperatures from $2-2.5$ GK. The photodisintegration path depends on the competition between charged particle and neutron emission. Due to the large uncertainty in the prediction of the low-energy $\alpha$ width, the path may not be well determined when the ($\gamma$,$\alpha$) and ($\gamma$,n) rates are within such a factor. At 2 GK, the ($\gamma$,$\alpha$) rate was almost a factor of 10 faster than the
($\gamma$,n) rate in \cite{rau06}, at 2.5 GK it was already lower by about 1/5. With the new rates from Table \ref{tab:reactivity}, the ($\gamma$,$\alpha$)/($\gamma$,n) rate ratio of $^{134}$Ce is 100 at 2 GK whereas the two rates are comparable at 2.5 GK. The difference comes from the newly calculated $^{134}$Ce($\gamma$,$\alpha$)$^{130}$Ba rate because the SMARAGD and NON-SMOKER predictions of the $^{134}$Ce($\gamma$,n)$^{133}$Ce rate are almost the same. The low-energy difference in the ($\gamma$,$\alpha$) rates comes from an improved numerical treatment of the charged-particle transmission coefficients at low energy in SMARAGD (see section 5.4.2 in \cite{tomreview} for a discussion). Thus, the path deflection at 2 GK is not shifted with respect to the one found in \cite{rau06} but a branching is introduced at 2.5 GK which removes some flux from the ($\gamma$,n) path leading to proton-richer Ce isotopes. Previously, the first path deflection point (where ($\gamma$,$\alpha$) dominated over ($\gamma$,n)) at 2.5 GK was at $^{132}$Ce. It may lose importance with the new rates because part of the photodisintegrations branch off already at $^{134}$Ce. This implies that the $\gamma$ process path would lie slightly closer to stability. It has to be cautioned, however, that the rates in the astrophysical energy window are still not well constrained by experiment.

As mentioned in Sec.\ \ref{sec:comparison} the deviation of the predictions from the data in Fig.\ \ref{fig:xs_ag} is not caused by a mispredicted $\alpha$ width but by problems in the predicted neutron and/or $\gamma$ width. Although this is inconsequential regarding the astrophysical $^{134}$Ce($\gamma$,$\alpha$)$^{130}$Ba photodisintegration reaction, it may affect the prediction of the ($\gamma$,$\alpha$)/($\gamma$,n) rate ratio discussed above. The $^{133}$Ce(n,$\gamma$)$^{134}$Ce rate (and its inverse $^{134}$Ce($\gamma$,n)$^{133}$Ce) mainly depends on the neutron width at 2 GK and is sensitive to both neutron and $\gamma$ width at temperatures 2.5 and 3 GK. Unfortunately, the dependences in the measured $^{130}$Ba($\alpha$,$\gamma$)$^{134}$Ce cross section and the inferred $^{133}$Ce(n,$\gamma$)$^{134}$Ce cross section are different. The overpredicted $^{130}$Ba($\alpha$,$\gamma$)$^{134}$Ce cross section can be due to a $\gamma$ width being too large or a neutron width being too small. On the other hand, a decreased $\gamma$ width would lead to a decrease in the predicted rate for $^{133}$Ce(n,$\gamma$)$^{134}$Ce at $T\leq 2.5$ GK while an increased neutron width leads to an increase in the same rate for $T<3$ GK. Without further information regarding the neutron and $\gamma$ widths in $^{134}$Ce it is not possible to draw further conclusions.

\begin{table*}
\caption{\label{tab:reactivity132}Stellar reactivities $R$ in cm$^3$s$^{-1}$mole$^{-1}$ and ground state contributions $X$ to the stellar rate as function of temperature $T$. Both quantities are given for the reactions $^{132}$Ba($\alpha$,$\gamma$)$^{136}$Ce and $^{136}$Ce($\gamma$,$\alpha$)$^{132}$Ba.}
\begin{ruledtabular}
\begin{tabular}{ddddd}
\multicolumn{1}{c}{$T$ (GK)}& \multicolumn{1}{c}{$R_{\alpha \gamma}$}& \multicolumn{1}{c}{$X_{\alpha \gamma}$}&\multicolumn{1}{c}{$R_{\gamma \alpha }$}& \multicolumn{1}{c}{$X_{\gamma \alpha }$}\\
\hline
  0.15 &  1.156\times 10^{-84} & 1.00 &   2.0268\times 10^{-90}   &    0.1822  \\
  0.20 &  1.085\times 10^{-72} & 1.00 &   2.0713\times 10^{-74}   &    0.1709  \\
  0.30 &  1.303\times 10^{-58} & 1.00 &   3.2317\times 10^{-56}   &    0.1425  \\
  0.40 &  7.324\times 10^{-50} & 1.00 &   2.3519\times 10^{-45}  &     0.1254  \\
  0.50 &  1.046\times 10^{-43} & 0.9999&  6.7055\times 10^{-38}  &     0.1136  \\
  0.60 &  4.611\times 10^{-39} & 0.9994&  2.2899\times 10^{-32}  &     0.1042  \\
  0.70 &  2.155\times 10^{-35} & 0.9977&  4.7843\times 10^{-28}  &     0.09549  \\
  0.80 &  2.138\times 10^{-32} & 0.9941&  1.5035\times 10^{-24}  &     0.08691  \\
  0.90 &  6.882\times 10^{-30} & 0.9876&  1.2147\times 10^{-21}  &     0.07838  \\
  1.00 &  9.487\times 10^{-28} & 0.9777&  3.5622\times 10^{-19}  &     0.07006  \\
  1.50 &  2.248\times 10^{-20} & 0.8766&  9.5831\times 10^{-11}  &     0.03618  \\
  2.00 &  6.715\times 10^{-16} & 0.7316&  1.1248\times 10^{-5}  &      0.01739  \\
  2.50 &  8.242\times 10^{-13} & 0.5934&  3.4109\times 10^{-2}  &      0.008395  \\
  3.00 &  1.642\times 10^{-10} & 0.4821&  1.3169\times 10^{1}  &       0.004211  \\
  3.50 &  9.888\times 10^{-9} &  0.3993&  1.3340\times 10^{3}  &       0.002275  \\
  4.00 &  2.457\times 10^{-7} &  0.3379&  5.0494\times 10^{4}  &       0.001349  \\
  4.50 &  3.043\times 10^{-6} &  0.2895&  8.6088\times 10^{5}  &       8.633\times 10^{-4}  \\
  5.00 &  2.151\times 10^{-5} &  0.2482&  7.4991\times 10^{6}  &       5.756\times 10^{-4}  \\
  6.00 &  3.090\times 10^{-4} &  0.1781&  1.3316\times 10^{8}  &       2.509\times 10^{-4}  \\
  7.00 &  1.534\times 10^{-3} &  0.1208&  7.9416\times 10^{8}  &       8.864\times 10^{-5}  \\
  8.00 &  4.270\times 10^{-3} &  0.07657& 2.7269\times 10^{9}  &       2.342\times 10^{-5}  \\
  9.00 &  8.507\times 10^{-3} &  0.04542& 6.6817\times 10^{9}  &       5.033\times 10^{-6}  \\
 10.00 &  1.358\times 10^{-2} &  0.02575& 1.2869\times 10^{10}  &      9.795\times 10^{-7}  \\
\end{tabular}
\end{ruledtabular}
\end{table*}

\begin{table*}
\caption{\label{tab:ratefit132}Fit parameters for the REACLIB fit format \cite{adndt} for $^{132}$Ba($\alpha$,$\gamma$)$^{136}$Ce and its inverse. Note that the value for $^{136}$Ce($\gamma$,$\alpha$)$^{132}$Ba obtained in this parameterization has to be multiplied by the ratio of the temperature-dependent partition functions $G(^{132}\mathrm{Ba})/G(^{136}\mathrm{Ce})$ \cite{adndt} to get the stellar reactivity.}
\begin{ruledtabular}
\begin{tabular}{lddddddd}
&\multicolumn{1}{c}{$a_0$}&\multicolumn{1}{c}{$a_1$}&\multicolumn{1}{c}{$a_2$}&\multicolumn{1}{c}{$a_3$}&\multicolumn{1}{c}{$a_4$}&\multicolumn{1}{c}{$a_5$}&\multicolumn{1}{c}{$a_6$}\\
\hline
($\alpha$,$\gamma$)&-1155.448&-121.1251&25.24542&1271.057&-87.73028&4.764612&-404.0757\\
($\gamma$,$\alpha$)&-1130.40071984&-126.44344235&25.24542&1271.057&-87.73028&4.764612&-402.5757
\end{tabular}
\end{ruledtabular}
\end{table*}

Similarly to the $^{130}$Ba($\alpha$,$\gamma$)$^{134}$Ce case discussed above, we also present the stellar reactivities for $^{132}$Ba($\alpha$,$\gamma$)$^{136}$Ce and its reverse rate in Table \ref{tab:reactivity132}. Again, the potential of \cite{McF} is used with the SMARAGD code, version 0.8.3s. The parameters for a fit of the reactivities in REACLIB format are given in Table \ref{tab:ratefit132}, along with the ground state contributions $X_{\alpha \gamma}$ and $X_{\gamma \alpha}$ to the stellar rate for the ($\alpha$,$\gamma$) and the ($\gamma$,$\alpha$) rate, respectively. The $X_{\gamma \alpha}$ for this reaction are comparable to the ones for $^{134}$Ce($\gamma$,$\alpha$)$^{130}$Ba.

The reaction $^{136}$Ce($\gamma$,$\alpha$)$^{132}$Ba was pointed out in \cite{rau06} because $^{136}$Ce could become a new deflection point in the photodisintegration path if this rate is increased. As explained above, this rate has to be compared to the $^{136}$Ce($\gamma$,n)$^{135}$Ce rate which is also only theoretically known. Again, the prediction of the ($\gamma$,n) rate remains almost unchanged with respect to the NON-SMOKER value. Therefore, the much higher new ($\gamma$,$\alpha$) rate exceeds the ($\gamma$,n) rate already at low plasma temperature and therefore the ($\gamma$,n) flow from stability would be interrupted already at this isotope. This renders the above discussed $^{134}$Ce deflection (or branching) unimportant and
forces the $\gamma$ process flow to move even closer to stability. It implies that the $p$ isotope $^{130}$Ba does not directly receive contributions from the Ce isotopic chain but only via ($\gamma$,n) reaction chains from neutron-richer Ba isotopes. The same cautionary remark as made above, however, also applies here. The validity of the potential by \cite{McF} has been proven by the experiment only at higher energy than required in the calculation of the astrophysical reaction rate.

In summary, the present experiment and its interpretation are exemplary for the typical difficulties encountered in the determination of charged-particle cross sections and reaction rates for astrophysics and, more specifically, the $p$ process. The cross sections at astrophysical energies are unmeasurably tiny. On the other hand, the cross sections at higher, experimentally accessible, energies show different sensitivities to nuclear properties than those at lower energy because more reaction channels are open and more particle widths contribute. A further fundamental difference is that charged-particle widths at astrophysically relevant energies tend to be much smaller than the $\gamma$ width, which is not the case at the usually investigated higher energies. Regarding $\alpha$ capture, the leading uncertainty in the astrophysical rate comes from the optical $\alpha$+nucleus potential. Measuring at higher energies, though, the various contributions and dependences of the particle widths and the $\gamma$ width have to be disentangled by combining data for different reactions. Regarding the $\alpha$ width, this can be achieved by (n,$\alpha$) reactions \cite{koeh}, but also by ($\alpha$,n) measurements which unfortunately have a strongly negative reaction $Q$ value in proton-rich nuclei. Therefore the optical potential can only be tested at comparatively high energies and theoretical considerations have to be invoked to arrive at an astrophysical rate. Nevertheless, ($\alpha$,n) measurements even close to the neutron emission threshold are scarce for this mass region. Further ($\alpha$,n) data are required and also further experimental efforts to access ($\alpha$,$\gamma$) cross sections at even lower $\alpha$ energy.

\begin{acknowledgments}

This work has been supported by the European Research Council StG. 203175 and OTKA K101328 and NN83261(EuroGENESIS). TR is supported by the European Commission within the FP7 ENSAR/THEXO project.

\end{acknowledgments}


\begin{thebibliography}{99}
\bibitem{talysrates} The TALYS rates were obtained directly from the code by using the built-in switch in the input file.
\end{thebibliography}
\end{document}